# The Emergence of Life as a First Order Phase Transition

Cole Mathis

*Beyond Center for Fundamental Concepts in Science*

*Department of Physics*

*Arizona State University, Tempe AZ*

Tanmoy Bhattacharya

*Sante Fe Institute, Sante Fe, NM*

*Los Alamos National Lab, Los Alamos, NM*

Sara Imari Walker

*Beyond Center for Fundamental Concepts in Science*

*School of Earth and Space Exploration*

*Arizona State University, Tempe AZ*

*Blue Marble Space Institute of Science, Seattle WA*

# Abstract

It is well known that life on Earth alters its environment over evolutionary and geological timescales. An important open question is whether this is a result of evolutionary optimization or a universal feature of life. In the latter case, the origin of life would be coincident with a shift in environmental conditions. Here we present a model for the emergence of life in which replicators are explicitly coupled to their environment through the recycling of a finite supply of resources. The model exhibits a dynamic, first-order phase transition from non-life to "life," where the life phase is distinguished by *selection* on replicators. We show that environmental coupling plays an important role in the dynamics of the transition. The transition corresponds to a redistribution of matter in replicators *and their environment*, driven by selection on replicators, exhibiting an explosive growth in diversity as replicators are selected. The transition is accurately tracked by the mutual information shared between replicators and their environment. In the absence of successfully repartitioning system resources, the transition fails to complete, leading to the possibility of many frustrated trials before life first emerges. Often, the replicators that initiate the transition are not those that are ultimately selected. The results are consistent with the view that life's propensity to shape its environment is indeed a universal feature of replicators, characteristic of the transition from non-life to life. We discuss the implications of these results for understanding life's emergence and evolutionary transitions more broadly.

**INTRODUCTION**

Life and the Earth system are tightly coupled [1]. A prominent example is the dramatic change in Earth's atmosphere due to the evolution of oxygenic photosynthesis [2]. An important open question is whether life's propensity to shape its environment is a *universal* feature of life, characteristic of the origin of life itself, or if the coupling between life and its environment observed in our biosphere is a product of evolutionary optimization that may perhaps evolve in some planetary contexts but not others. One of the most distinctive features of life is replication—the ability to make copies—which engenders living matter with the capacity to sustain stable patterns of non-equilibrium behavior. Accordingly, numerous theoretical studies for the emergence of life have focused on the appearance of the first replicators, including identifying the conditions under which replicators can be selected from a prebiotic milieu [3–9]. Here we study a computational model for the emergence of replicating polymers, which includes coupling to an environment through recycling of a finite supply of resources, to address the role of the *environment* in driving the transition from non-living to living matter. We demonstrate that a key property of selection for replication in prebiotic systems under resource limited conditions is the feedback between replicators and their environment.

To study the role of environmental coupling, we consider a model prebiotic "replicative chemistry" with a finite supply of monomers, which must be recycled through polymer degradation to replenish resources available for synthesis and replication of polymers. By contrast, the majority of theoretical models for the emergence of replicators thus far have implemented reactor flows with a constant flux of monomers into the system and removal of chemical species via dilution, and therefore do not explicitly include feedback from the environment. Examples include the transition from "pre-life" to "life" as studied by Nowak and collaborators, where "pre-life" is defined as a generative chemistry with no replication (polymerization only), to be contrasted with "life", where replicators are selected [3–5]. In their model, a transition from "pre-life" to "life" is observed by externally modulating the replication rate of polymers: above a critical rate constant for replication, replicating polymers can be selected. Similar features have been noted by Wu *et al.* [6, 7] and Szathmary and Maynard Smith [8]. By explicitly coupling replicators to their environment, we show this transition can occur *spontaneously*, without tuning any relevant rate parameters externally, and is dynamically driven by the environment. We also demonstrate that the abrupt nature of the observed transition shares many features in common with first-order phase transitions as characterized in the physical sciences. Here the two phases, which we nominally call "nonlife" and "life," are distinguished, not by the discovery of replication, but by the absence and presence of *selection* for replication [1]. The dynamics observed demonstrate many of the hallmarks of *dynamic kinetic stability* (DKS) [11], where the life-phase is characterized by the kinetically driven stability of self-replication. We discuss the implications of these results for furthering our understanding the emergence of life and evolutionary transitions more broadly.

---

[1] We realize that defining life is a much more complex issue than simply defining life as the ability to replicate and use the terminology ``life'' heuristically to indicate a distinction between the two phases relevant to a key property of life, that is its ability to make copies.

**METHODS**

**Model Description**

We model the emergence and dynamics of replicators in an artificial prebiotic 'chemistry' consisting of two monomer types denoted by '0' and '1'. The properties of our model chemistry are fully specified by the rate constants $k_p$, $k_d$ and $k_r$ for polymerization, degradation and replication, respectively, the finite constant abundances of the two monomeric components '0' and '1', a constant $r$ specifying the minimal length of replicating sequences and fitness landscapes associated with sequence-specific replication and stability introduced below.

Polymerization occurs via addition of monomers to the end of growing sequences. Polymers can degrade into shorter sequences, which can occur at any bond within a given sequence with equal probability. To simplify the computational model, we adopt a common approximation in models for prebiotic polymerization that the inverse process of two short but non-monomeric sequences ligating to produce a longer polymer is sufficiently rare to be neglected (which would be the case, for example if monomers are much more common than dimers) [3, 4, 9]. All sequences of length $L \geq r$ can self-replicate such that polymers must be sufficiently 'complex' to copy themselves [12–14]. In this study, we set $r = 7$, such that the appearance of the first replicators is rare, but not so rare that we never observe it [6]. Changing $r$ changes the relative timescale of the transition, but does not qualitatively effect the results presented herein.

Since we are interested in the dynamics of replication in this work, and specifically the origins of life, we do not include the effects of mutation, which is well known to play an important role in evolution once life has already emerged [15], but is not expected to alter the qualitative features of the transition in the simplified model reported here. Therefore in our model, replication only functions to copy extant sequences, and does not produce novelty. Polymerization, however, does produce new sequences, and in our model, novelty is solely introduced through the prebiotic recycling of monomers via degradation and polymerization.

Simulations were implemented using a kinetic Monte Carlo algorithm [16, 17]. For more detailed discussion of the implementation of that algorithm in prebiotic recycling chemistries we refer the reader to [9] or [18]. In what follows, the polymerization, degradation, and replication rate constants were set to $k_p = 0.0005$, $k_d = 0.5000$, and $k_r = 0.0050$ respectively, and the system was initialized with 500 monomers each of '0' and '1' with no polymers present, unless otherwise noted. Since the reaction network is a closed mass system, the initial conditions specify the bulk composition of the system for all time.

**Two Fitness Landscapes: Static and Dynamic**

To explicitly couple the properties of replicators to that of their environment, we model the fitness of replicators as determined by two factors:

- A *static* fitness associated with a trade-off between stability and replicative efficiency that is an intrinsic 'chemical' property of individual polymer sequences.

- A *dynamic* fitness associated with resource availability in the environment.

The former concept of *static* fitness encompasses the components of the fitness landscape associated with the properties of specific polymer sequences, without taking any account of the availability of resources. We choose to model selection for replication and stability as a trade-off since in many real-world chemical systems molecules that fold well are typically not good self-replicators and conversely good self-replicators often do not fold well and are thus less resistant to degradation [19]. The latter concept of *dynamic,* environmentally-dictated, fitness, accounts for the availability of resources (free monomers) in the system, and is a unique feature of the resource-dependent replication model presented here (see also [9] or [18]).

*Static Fitness.* An important question in any model for the emergence of life is how sensitive the observed dynamics are to model parameters, or to the fitness landscape imposed in the absence of empirical data. In what follows, we therefore consider several static fitness landscapes, which vary in how well the composition of the 'fittest' sequence(s) matches the resource availability in the bulk environment. This allows us to determine if it is selection on replicators generally, or features specific to a particular fitness landscape (and its relationship to abiotic resource distributions), that drive the dynamics observed. We consider four cases:

- *Landscape I (LI): Replicators and stable sequences are rare*. For the first example, replicative efficiency increases with the number of '0' monomers in a sequence, and stability with the number of '1' monomers, such that homogeneous all-'0' sequences with $L \geq 7$ are the best replicators, and all-'1' sequences are the most stable. Since the bulk composition of the prebiotic environment consists of an equal number of '0' and '1' monomers in our simulations and polymerization does not favor any specific bond type, good replicators and stable sequences are very rarely produced abiotically and their composition does not reflect the abiotic distribution of resources.

- *Landscape II (LII): Replicators are rare, stable sequences are common*. For the second example, replicative efficiency increases with the number of '00' bonds in a given sequence, while '01' bonds increase stability. This allows replicators to be rare as determined by the rate of spontaneous polymerization, whereas stable sequences are more readily produced. Even in this case, the composition of the best replicators does not match the bulk composition of the environment, but the composition of the most stable sequences now does.

- *Landscape III (LIII): Replicators and stable sequences are common*. For the third example, replicative efficiency increases with the number of '10' bonds in a given sequence, while '01' bonds increase stability. For this fitness landscape, both efficient replicators and stable sequences are readily produced via spontaneous polymerization processes, and both the best replicators and most stable sequences reflect the composition of the environment.

- *Landscape IV (LIV ): All sequences with $L \geq 7$ replicate with equal efficiency*. For the final example, all sequences of length $L \geq 7$ replicate with equal efficiency (as would occur if replication were environmentally driven [9]). We use this as a control to determine if the features observed are intrinsic to a selection for *replicative fitness* or a more general property of selection for *replication* and the transition from prebiotic polymerization processes to more 'life-like' template-based replication.

For L*I*–L*III* the mathematical form for the trade-off between replication and stability is quantified as:

$$f(n) = 0.5 + \frac{n^2}{2(10+n^2)}, \tag{1}$$

following Szabo *et al.* [19], who implemented a similar trade-off among attributes of replicating polymers. In our implementation, the replicative fitness of a sequence $x_i$ with length $L$ is quantified by scaling its replication rate $k_r$ by a parameter $\alpha_r(x_i) = 1 + f(n)$ (which is a number between 1.5 and 2), where $n$ is the quantity which confers replicative efficiency (*i.e.,* '0' monomers, '00' bonds, and '10' bonds in L*I*–L*III*, respectively). Similarly, the stability of sequence $x_i$ is determined by scaling the degradation rate $k_d$ by the parameter $\alpha_s(x_i) = 1 - f(m)$ (which is a number between 0 and 0.5), where $m$ is the quantity which confers stability (*i.e.,* '1' monomers for L*I*, '01' bonds for L*II* and L*III*). Sequences with $L < 7$ do not replicate, so only the stability landscape is relevant for short sequences. In each example, this establishes a fitness landscape *intrinsic* to a polymer's specific sequence that is fixed within a given environmental context.

*Dynamic Fitness.* Since we are interested the coupling between replicators and their environment, we also introduce an *extrinsic*, dynamic term to replicative efficiency, which is determined by the availability of free monomers in the environment. To this end, the replication rate for sequence $x_i$ is weighted by a factor $\beta(x_i) = \Sigma_{n_i}^{L-1} y_{n_i} y_{n_i+1}$, where $y_n$ is the abundance of the monomer species at position $n$ in sequence $x_i$. This term yields a computationally tractable resource-dependent replication rate that is also sequence dependent. This term may be motivated as a sum over all possible nucleation events on a template (see *e.g.,* [18] for an explicit example as applies to ribozyme recycling). As such, the replication rate of a given sequence $x_i$ depends in part on how well its sequence composition matches the relative abundances of '0' and '1' monomers in the environment. Since the abundances of '0' and '1' monomers change over time as monomers are consumed via polymerization and replication and generated via degradation, this creates a dynamic, environmentally dictated fitness landscape that is a central feature of any resource-constrained dynamics. We expect qualitative features of the dynamics observed here to be a general feature of species-specific and resource-dependent replication, independent of the particular functional form of $β(x_i)$.

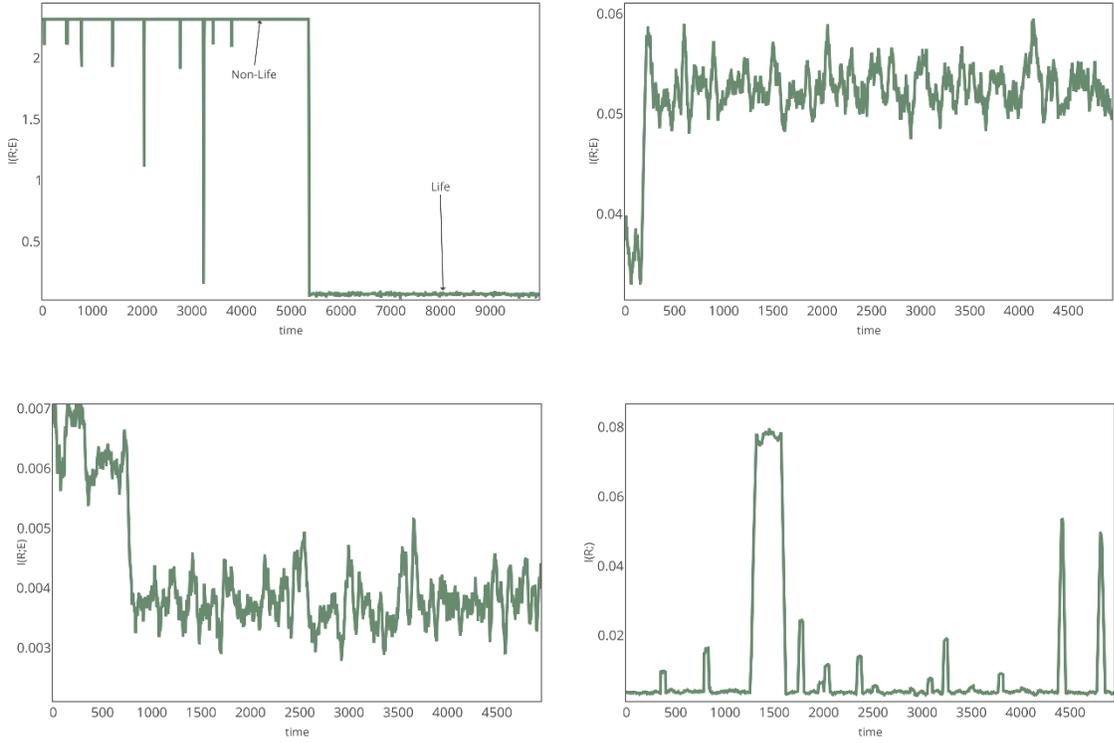

FIG. 1. Typical time series of the mutual Information shared between replicators and their environment (free monomers) for each fitness landscape. Top left: L$I$, Top right: L$II$, Bottom left: L$III$, Bottom right: L$IV$. The phase transition is clearly evident in the abrupt shift(s) in I($R;E$) observable in each case.

**Tracking the Selection of Replicators with Mutual Information.**

To characterize the dynamics of the observed phase transition, we employ mutual information, a common tool in information theory, which measures the mutual dependence of two variables within a dynamic time series by quantifying how much information the two variables share in common. We use mutual information, I, to measure the extent to which the composition of replicators is determined by their environment, and vice-versa. We define the sets $R$ and $E$ which contain ordered pairs that track the number of '0' and '1' monomers in replicators ($R$) and in free monomers in the environment ($E$), allowing us to measure I($R : E$), defined as the mutual information shared between replicators and their environment. See Supplementary Information for additional details.

**RESULTS**

**A First Order Phase Transition from Non-life to Life**

Two long-lived states are observed, which we nominally call "non-life" and "life". These two phases are dominated by polymer formation via polymerization or via replication, respectively. While the non-life phase here shares features in common with "pre-life" as previously characterized [3], it also has some striking differences—we therefore use "non-life" rather than "pre-life" (in our system many transitions fail to complete so the emergence of replicators and life is not inevitable, thus "non-life" is more appropriate).

For fixed values of $k_p$, $k_d$ and $k_r$, the system exhibits a spontaneous and abrupt transition from non-life to life for all four fitness landscapes investigated. The transition is accurately tracked by measuring the mutual information I($R;E$) between the composition of extant replicators and free monomer resources, as shown in Fig. 1. The details of the transition are dependent on the static fitness landscape chosen, but the transition exists *independent* of the nature of the landscape chosen. For L*I* (Fig. 1, top left), mutual information *decreases* through the transition, since the composition of selected replicators differs from that of their environment. This is true also of L*II* (Fig. 1, top right), however the magnitude of the difference in I($R;E$) between the two phases is less dramatic, due to selection of stable, less efficiently replicating sequences whose composition does match bulk resources, in addition to efficient replicators whose composition does not. For L*III*(Fig. 1, bottom left) mutual information *increases* since the composition of selected replicators reflects the bulk composition of their environment. For L*IV* (Fig. 1, bottom right), where there is no selection on specific sequences, the transition is visible in the rapid changes in the mutual information between a state of high and a state of low mutual information as stochastically determined replicators are selected and accumulate through exponential growth, till a large fluctuation leads to a different replicator dominating. For L*IV* the transition is frustrated for the particular choice of parameters: sequences are transiently selected, but the system settles back before the sequences dominate. The absolute value of I($R$;$E$) and the size of the fluctuations is sensitive to the fitness landscape imposed, however, the presence of an abrupt, spontaneous transition is apparent in each case, and accurately tracks the on-set of selection on the properties of replicators.

Since the transition is apparent for each landscape explored, we focus on L$I$, as we expect this landscape to exhibit the most interesting dynamics, given that replicators and stable sequences are rare and not readily produced prebiotically. We also expect this case to be the most realistic as functional biopolymers are sparse in sequence space and often do not share their composition with the ambient environment.

The dynamics of the transition from non-life to life displays many hallmarks of physical first-order phase transitions. The result of a Kolmogorov-Smirnov (KS) test suggest that the distribution of wait times until the transition successfully completes is consistent with the expected exponential distribution (left, Fig. 2) [1]. The frequency that the transition occurs is dependent on both the composition of replicators and the environment (right, Fig. 2). This is distinct from other models that do not account for environmental feedback [3, 6]—here the transition is *not* coincident with the first 'discovery' of a sequence capable of replication, since replicators can exist in non-life. Instead, the transition occurs when replicators and the environment share similar resource distributions (high extrinsic fitness). Since both monomer species are equally abundant in the initial distribution of resources for the examples reported here, the nucleation event is typically mediated by heterogeneous replicator(s) composed of a roughly equal number of '0's and '1's (this is true for all four landscapes). These are not the sequences that are ultimately selected in the life phase for L*I*, which include only the homogeneous, fit sequences. Given that the transition is spontaneous, abrupt and exponentially distributed we consider the dynamics to be indicative of a genuine first-order phase transition. We note that similar to other first-order transitions, there are often many frustrated transitions prior to a successful phase transition (see Fig. 1, top left), which can occur when lack of selection on fit sequences leads to a failure in the transition to run to completion (see also Fig. 1, bottom right).

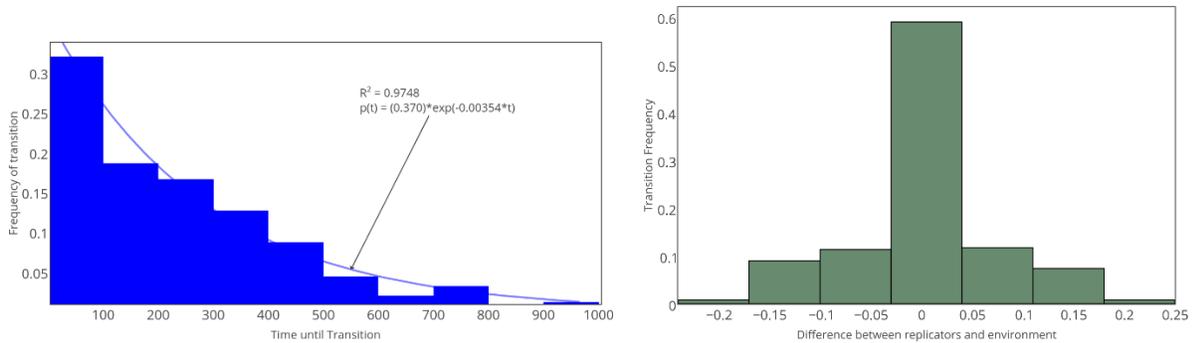

FIG. 2. Left: The distribution of waiting times until the phase transition occurs for L$I$ is shown. It follows an exponential distribution, indicative of a first order phase transition due to large fluctuations. Right: The frequency of successful phase transitions for L$I$ as a function of the difference between the composition of extant replicators and their environment. Data shown is for an ensemble statistic of 256 simulations. Simulation parameters were set as $k_p$ = 0.0005, $k_d$ = 0.5000, and $k_r$ = 0.0050 for both figures.

**The Dynamics of the Transition from Non-life to Life**

In the non-life phase, long sequences are exponentially rare, and the majority of system mass is in monomers and dimers (not shown). Sequences of all lengths have relatively similar composition, as shown in the top panel of Fig. 3. The composition of extant polymers is
reflective of the combined effects of the abiotic availability of resources, and, for L*I*–L*III*, the stability landscape established by Eq. 1. Replicators can exist in the non-life phase, albeit at exponentially low abundance. These typically have compositions reflective of the
abiotic distribution of resources and form via polymerization.

In the life phase, the composition of replicators need not reflect the bulk composition of the environment; instead, replicator composition is determined by *selection* of the fittest sequences. This can in turn lead to restructuring of the distribution of resources in the en-vironment, as shown in Fig. 3 for L$I$, where selected replicators are primarily homogeneous '1's or '0's. Due to resource constraints, selection on replicators drives a transition in the composition of shorter sequences. In the life phase, short sequences obtain the opposite compositional signature to that of replicators (bottom panel, Fig. 3). The compositional reversal is seen only below $L = 6$. Although $L = 6$ sequences cannot replicate, they are formed primarily via degradation of $L = 7$ replicators and thus their formation is dominated by self-replication (via formation, then degradation of $L = 7$ sequences). In the life phase, replicators are selected based on their *intrinsic* fitness and not strictly how well their composition matches the environment. The defining feature of the life phase is therefore not necessarily the presence of replicators, which exist in both phases. Instead, the defining characteristic of "life" in this model is that the distribution of resources is dictated by *selection* on the properties of replicators, and that selection only operates in the life phase.

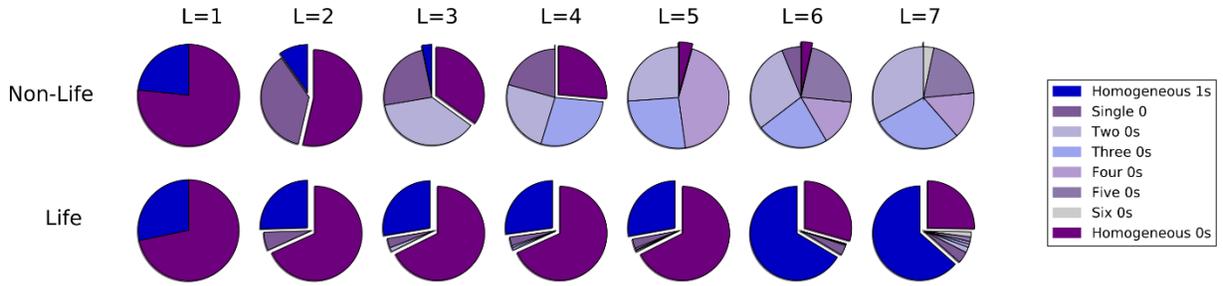

FIG. 3. Ensemble averaged compositions of all sequences with $L \leq 7$ for L$I$. The distributions in the top-panel characterize the non-life phase (no selection on replicators), and the bottom-panel characterize the life phase (selection on replicators). Data is averaged over 100 simulations and simulation parameters are $k_p = 0.0005, k_d = 0.5000,$ and $k_r = 0.0050$

Fig. 3 demonstrates that, due to resource constraints, the selection of replicators can coincide with dynamic restructuring of the entire system (including both monomer and nonreplicating ($L < 7$) sequence populations). Fig. 4 shows an ensemble averaged phase space trajectory through this restructuring for L$I$. The phase transition moving from non-life to life phases is highly unstable and dominated by degradation. In both the non-life and life phases, polymer formation rates (polymerization and replication) balance rates for polymer degradation, with ratios of formation/degradation $\sim 1$. However, the life and non-life phases are clearly distinguished in phase space by very different values for the mutual information (here I($R;E$) ~ 3.0 for non-life and I($R;E$) ~ 0.25 for life, for results in Fig. 4).

The rampant degradation observed through the phase transition results in a rapid and dramatic restructuring of the extant polymer population and a steep slope in the rate of sequence exploration, as observed in Fig. 5. This is characteristic of the phase transition from non-life to life, independent of the replicative fitness landscape. This restructuring arises as a result of reallocation of mass from shorter sequences to replicators, which must occur via degradation to monomers that can then be consumed via replication. The extant

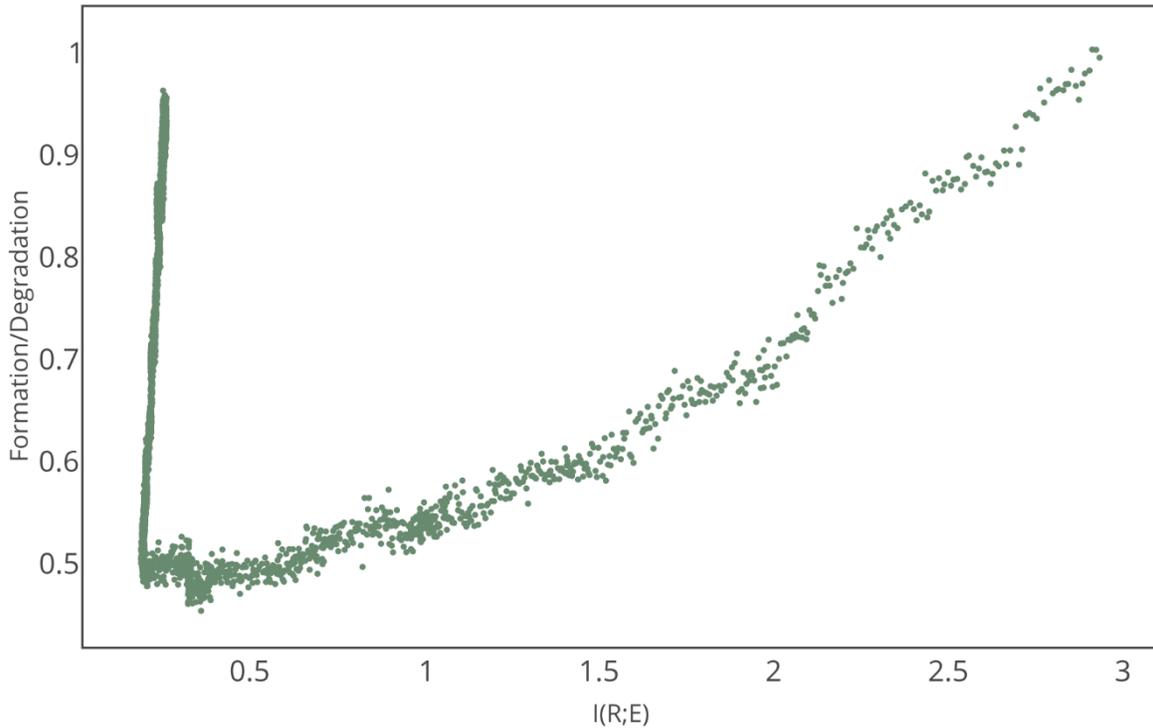

FIG. 4. Phase trajectory for an ensemble of 100 systems transitioning from non-life to life (plotted versus time the system would move from left to right). Axes are the mutual information I($R;E$) between replicators and environment ($x$ axis) and the ratio of formation (polymerization and replication) to degradation rates ($y$ axis). Simulation parameters are $k_p$ = 0.0005, $k_d$ = 0.5000, and $k_r$ = 0.0050

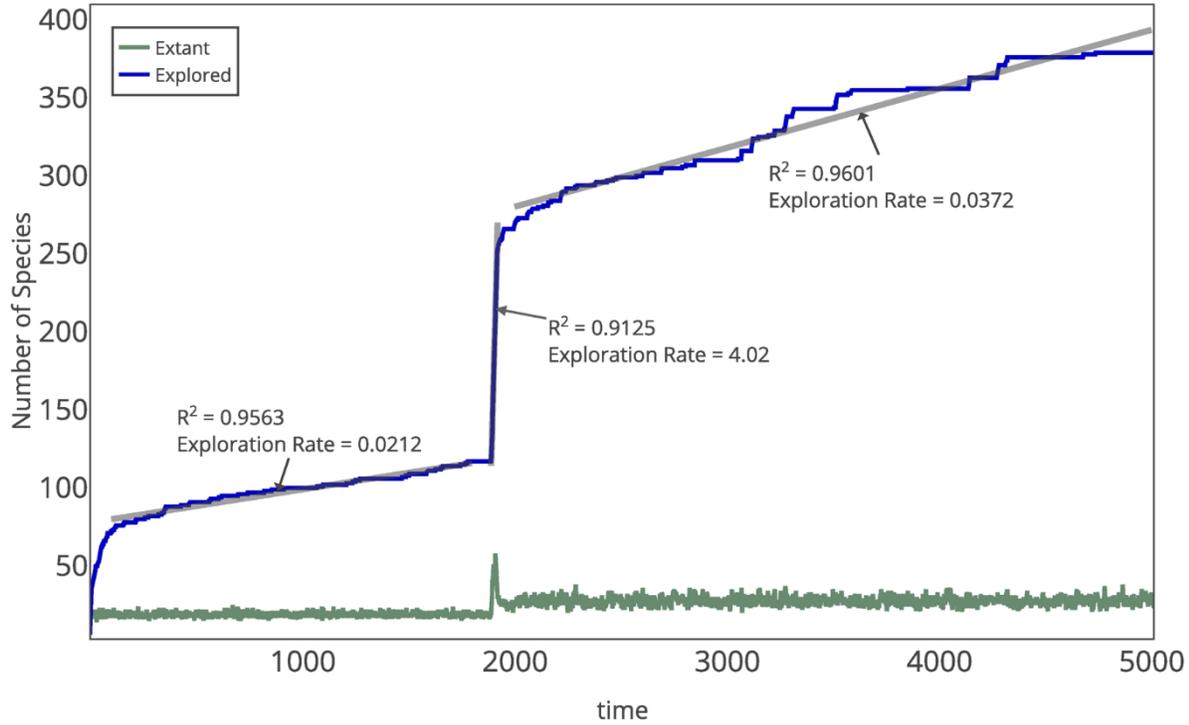

FIG. 5. Exemplary time series for the extant species population size and total number of sequences explored by the system. Linear fits to the explored species are shown. The exploration rate is 75% faster during the life phase compared to the non- life phase, and is 2 orders of magnitude larger during the transition. Simulation parameters are $k_p$ = 0.0005, $k_d$ = 0.5000, and $k_r$ = 0.0050

diversity and the rate of introduction of new sequences are both higher in the life phase than the non-life phase (Fig. 5), attributable to the higher turnover rate of resources in the life phase (due to the higher assembly rate of polymers via replication).

Shown in Fig. 6 is an example time series for the evolution of all sequences with $L = 7$ for L$I$, binned by sequence composition, for a set of parameters where the transition is prolonged enough to resolve details of the restructuring. Resources constraints enforce selection of sequences in complementary pairs that maintain the symmetry of the bulk resource distribution of the environment (50% '0's and 50% '1's). The system subsequently undergoes a series of abrupt transitions associated with increasing sequence homogeneity, where replicator composition increasingly departs from that of the bulk environment.

**The Timescale for Life's Emergence**

The phase transition from non-life to life described here is a robust feature of the dynamics, observed for different fitness landscapes with qualitatively similar features. Here we demonstrate that the observed features are also qualitatively robust over a large range of parameters values. Quantitative differences arise in the final abundances of replicators and in the timescale for the transition to occur, which are both sensitive to the specific details of the prebiotic chemistry under consideration. Fig. 7 shows the average time to complete the phase transition as a function of the degradation and replication rate constants, $k_d$ and $k_r$ for L$I$. For the results presented, the transition was identified as complete when 75% of the total replicating mass was allocated in homogeneous (fit) sequences.

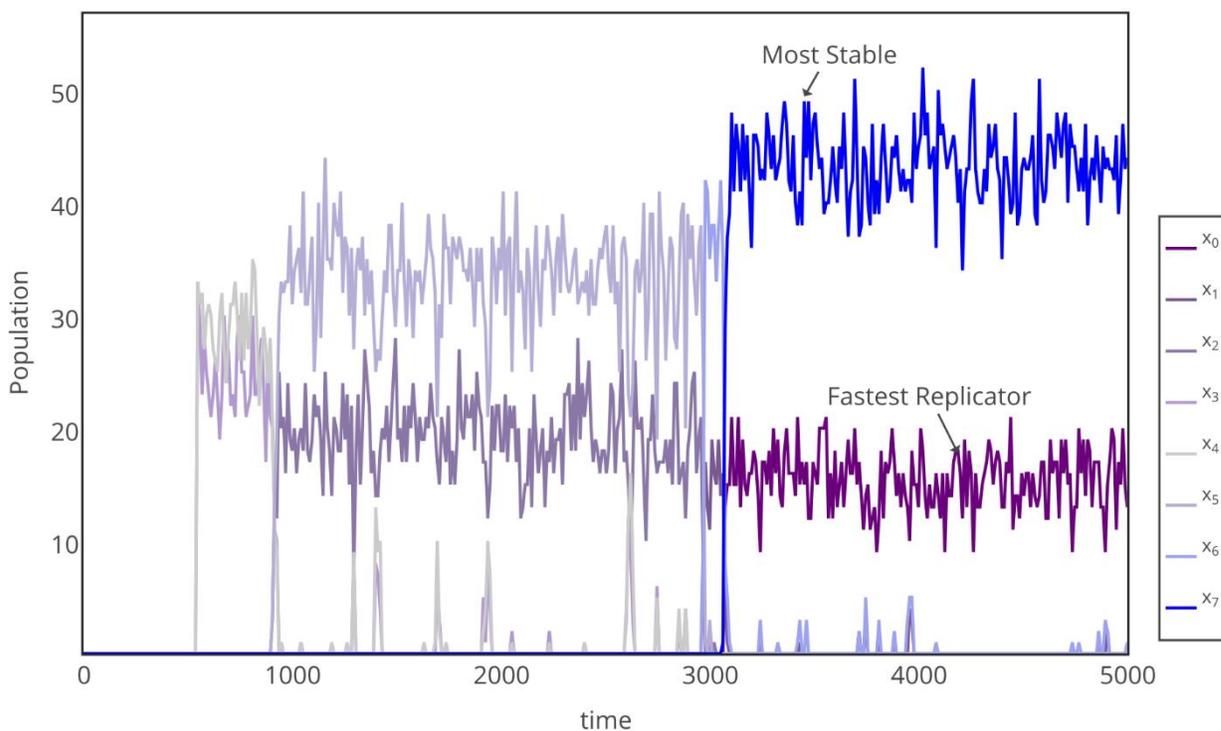

FIG. 6. Series of transitions in the selection of fit, homogeneous '0' and '1' length $L = 7$ replicators for L$I$. Here, the subscript denotes the number of '1' monomers in the sequence (*e.g.* $x_0$ contains no '1's, $x_1$ bins all polymers with a single '1' monomer, and $x_7$ contains all '1's). Simulation parameters are $k_p = 0.0005, k_d = 0.9000,$ and $k_r = 1.000$.

One might *a priori* expect the transition to be most rapid (favored) for fast replication (high $k_r$) and slow degradation (low $k_d$), however this is not always observed. For high degradation rate $k_d$ = 5.0, the time to the transition is largely independent of $k_r$ (Fig. 7). Lowering the degradation rate ($k_d$ = 1.0 and $k_d$ = 0.5 Fig. 7) increases the dependence of the transition time on $k_r$, which, on average, occurs most rapidly for relatively low $k_r$. This counterintuitive behavior arises as a result of the resource constraints. For high degradation rates, there is a high rate of turnover increasing the likelihood of discovering functionally fit sequences, but the probability of survival is low, so the transition time is long regardless of replicative efficiency. For lower degradation rates, high replication rates lock resources in less fit sequences, frustrating the system's restructuring, also leading to long transition times.

The rate of degradative recycling seems to be the primary factor in determining the transition timescale. Fig. 8 shows the transition time observed for different abiotic resources abundances, quantified by the ratio $R$ of the total number of '1' monomers to total system mass. The transition timescale is not expected to be symmetric with respect to the relative abundance of '0' and '1' monomers. For large values of $R$ (environments rich in '0' monomers that confer stability), where recycling is inherently slower, the average transition time may be much longer than in environments with fewer stable polymers. Our data supports this expectation although the variation in transition times is large. These features suggest that environments which engender degradative recycling at a moderate rate may be the most conducive to nucleating the origin of life under resource-limited conditions.

**DISCUSSION**

We have demonstrated the existence of a spontaneous transition from non-life to life, which arises due to explicit incorporation of environmental feedback and displays many

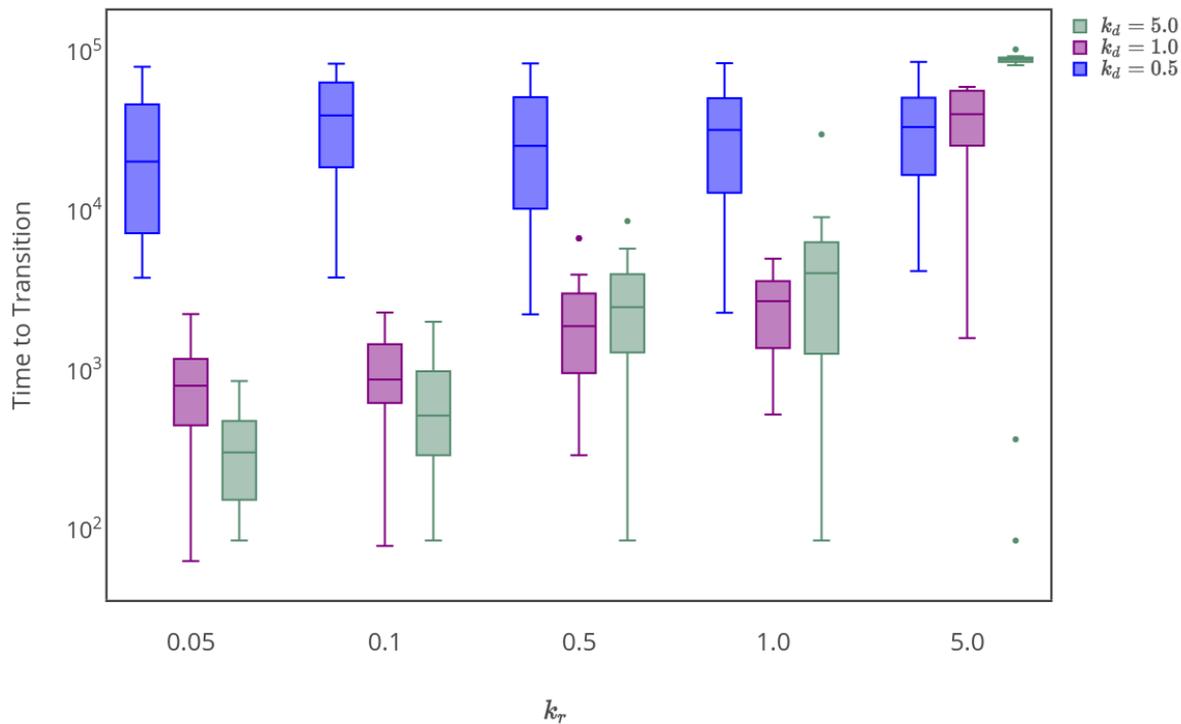

FIG. 7. Timescale for completing the phase transition as a function of reaction rate constants for replication $k_r$. Data from 25 simulations is shown, all data points are included in the box and whisker plots. The center line for each distribution is the median, the boxes contain half the data points and the bars show the range. Three values of the degradation rate constant $k_d$ are shown, 0.5 (blue), 1.0 (purple), 5.0 (green), the polymerization rate constant was fixed at $k_p$ = 0.0005.

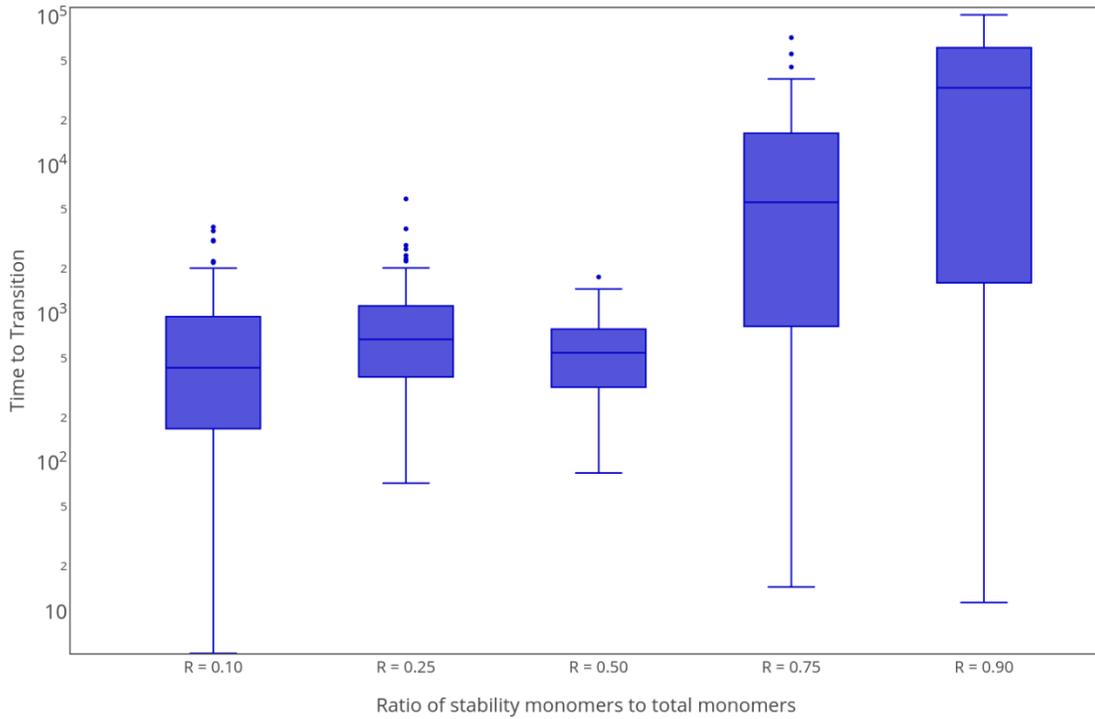

FIG. 8. Timescale for completing the phase transition as a function of the abiotic distribution of resources. Here, the parameter $R$ is the ratio of '1' monomers (which confer stability) to total system mass. Data from over 100 simulations is shown, all data points are included in the box and whisker plots. The center line for each distribution is the median, the boxes contain half the data points and the bars show the range. Parameters: $k_p$ = 0.0005, $k_d$ = 0.5000, and $k_r$ = 0.0050

features in common with first order physical phase transitions. It might be argued that the dynamics reported here do not represent a true phase transition. In the study of equilibrium physical systems, free energy is the quantity which is minimized to determine the state of the system [20]. Typically, this involves a play-off between minimizing total energy and maximizing entropy. When these two favor different results, a system is expected to exhibit a first order phase transition from order to disorder. Here, in our dynamical scenario, a similar tradeoff happens between two processes that consume and try to minimize the number of free monomers (which may be related to the minimization of free energy [21]). These two different ways—*viz.,* maximizing the number of bonds via polymerization, or by maximizing the number of polymers via replication—yield distinct results with a sharp boundary between them, which motivates the classification of the observed dynamics as a phase transition. Future work will detail whether this is merely a useful analogy or indicative of a deeper connection.

Importantly, the most distinguishing feature of the life phase in our model is not the presence of replicators, since these can also exist in non-life. Instead it is *selection* on the properties of replicators (such as replicative efficiency and stability in the examples presented here). Selection in turn necessitates a redistribution of matter due to limited resource availability. This restructuring is coincident with a sharp transition in the mutual information shared by replicators and their environment. Previous work connecting information theory to life's origins reported that the probability to discover a self-replicator by chance should depend exponentially on the availability of monomers composing it [22]. Our results demonstrate an additional necessary feature: in the case of resource constrained replication, replicators and environment share a similar composition (*e.g.,* have high mutual information). This enables exponential growth of the replicator population based on high dynamic fitness, which in turn enables selection on the properties of new replicators discovered. When the fittest replicators do not match the bulk composition of the system, they force a redistribution of resources to accommodate their selection. We further note that very few measures have been proposed to explicitly quantify the origin of life transition. Here, mutual information between replicators and environment accurately measures the progress of the phase transition reported (perhaps acting as an order parameter), independent of the specific attributes of the replicator selected. Future work should elucidate the relationship between the fitness landscape and the system dynamics and magnitude of the mutual information. This will help to identify how broadly applicable this approach is, and perhaps provide insights to other candidate scenarios for the origin of life, such as in the formation of autocatalytic sets [**?** ].

While we have nominally identified selection on replicators with "life" in this simple model, we note that the presence of replication is perhaps a necessary but not sufficient criterion to define life (see *e.g.*, [23]), which remains an important open philosophical and scientific question [24]. The information-theoretic characterization of this transition is consistent with proposals that life is most defined by its informational properties [22, 23] (here, replicators might be interpreted as driving the dynamics of the entire system in a "top-down" manner due to adaptive selection [25]). The life phase may be interpreted as a state where the kinetics of individual replicators (*e.g.,* as quantified by their replicative efficiency and stability) dictate the behavior of the entire system, consistent with the notion that life is a kinetically driven state of matter [11]. Although our motivation is to understand the origin of life utilizing this model system, we note that the model is sufficiently general to capture features that may be universal to a broader class of evolutionary transitions. In particular, the dynamics could be universally characteristic of the discovery of novel, selectable patterns in the distribution of resources among replicating populations. For example, the abrupt nature of the transition shares features in common with punctuated equilibrium [26]. The dynamics of this phase transition also demonstrate behavior that may be characteristic to niche construction and/or mass extinctions. In particular, the system's restructuring necessitates a period of instability driven by rampant destruction of extant diversity (extinction), which is followed by an explosion in novel diversity. The relationship to the phase transition reported here could be tested, for example, by analyzing the connection between resource distribution patterns and abrupt evolutionary transitions in the evolutionary record of life on Earth.

Interestingly, the features most characteristic of the phase transition reported are heavily dependent on degradative recycling of finite resources, which mediates selection on fit sequences by recycling less fit ones and yields an abrupt transition due to rapid resource re-allocation. This suggests new perspectives regarding the role of degradation in the origin of life, which is typically viewed as an impediment in prebiotic chemistry, rather than a process central to early evolution [27]. Cast under new light in the resource-constrained dynamics observed here, it is perhaps not a coincidence that RNA, as a biopolymer that played a prominent role in early evolution, is highly susceptible to hydrolysis, perhaps resolving an apparent paradox in the origin of life [28]. The properties of this phase transition are in principle testable in the laboratory in experimental systems that permit recycling of biopolymers, for example as reported in [18]. In particular, the observed dynamics should place further constraints on the kinds of chemistries (defined by relative rates $k_d$, $k_r$ and $k_p$ and relative resources abundances) that are most conducive to mediating the transition from non-life to life (see also *e.g.,* [29]).

Due to the explicit coupling between replicators and environment, the transition reported here displays many features one might expect for a newly emergent biosphere that are not observable in open-flow reactor models. In particular, restructuring during the phase transition drives a vast increase in extant diversity and in the rate of exploration of novel diversity. This indicates that the emergence of life should coincide with an explosive growth of novelty in resource limited systems. Concomitantly, during the transition the system is dramatically restructured, indicating that the emergence of life should have significantly altered the environment of the early Earth. It is well known that biology alters its environment over evolutionary and geological timescales, and that the presence of life defines many features of the Earth-system. Our results indicate that this may be a universal characteristic of life, from the very first appearance of replicators and is most dramatic in cases where life is composed of sequences rarely produced abiotically.

The model includes the possibility of many frustrated trials before life first emerged (see *e.g.,* Fig. 2), with success entailing a transformation of the environment as a necessary component of the process of biogenesis (perhaps consistent with the notion of a "Gaian bottleneck" [30]). These features indicate that it should be difficult to retrace the precise history of the origin of life: the replicators that are ultimately selected will, in general, neither be reflective of the ancestral planetary environment from which life first emerged, nor will they be representative of the replicators that first nucleated the origin of life. Thus, as is often suggested, here we see an explicit example that the conditions favoring the emergence of life may not be the same as those favoring its subsequent evolution.

Finally, we point out that simple replicators such as those presented here may not be the most effective architecture for a self-reproducing system. In this model, the total composition of the system remains fixed, what life does is restructure the distribution of matter within the system, due to the propagation of selectable replicating resource allocation patterns. An interesting open question is how this phase transition might play out for more life-like replicative systems, such as those with the architecture of a von Neumann self-reproducing automata [23, 31, 32], a subject we leave to future work.

This project/publication was made possible through support of a grant from Templeton World Charity Foundation. The opinions expressed in this publication are those of the author(s) and do not necessarily reflect the views of Templeton World Charity Foundation. The authors wish to thank Paul C.W. Davies and Nigel Goldenfeld for constructive conversations on this work and the Aspen Center for Physics (supported in part by the National Science Foundation under grant no. PHY-1066293) for hosting SIW and TB, where the initial seeds of the idea that nucleated this project were matched with the right environment.

[1] E. Smith and H. J. Morowitz, *The Origin and Nature of Life on Earth: The Emergence of the Fourth Geosphere* (Cambridge University Press, 2016).

[2] B. E. Schirrmeister, M. Gugger, and P. C. Donoghue, Palaeontology **58**, 769 (2015).

[3] M. A. Nowak and H. Ohtsuki, Proceedings of the National Academy of Sciences **105**, 14924 (2008).

[4] M. Manapat, H. Ohtsuki, R. Bürger, and M. A. Nowak, Journal of theoretical biology **256**, 586 (2009).

[5] H. Ohtsuki and M. A. Nowak, Proceedings of the Royal Society B: Biological Sciences **276**, 3783 (2009).

[6] M. Wu and P. G. Higgs, Journal of Molecular Evolution **69**, 541 (2009).

[7] M. Wu and P. G. Higgs, Biology direct **7**, 42 (2012).

[8] E. Szathmáry and J. Maynard Smith, Journal of Theoretical Biology **187**, 555 (1997).

[9] S. I. Walker, M. A. Grover, and N. V. Hud, PLoS ONE **7**, e34166 (2012).

[10] We realize that defining life is a much more complex issue than simply defining life as the ability to replicate and use the terminology "life" heuristically to indicate a distinction between the two phases relevant to a key property of life, that is its ability to make copies.

[11] A. Pross, Origins of Life and Evolution of Biospheres **35**, 151 (2005).

[12] G. von Kiedrowski, Angewandte Chemie International Edition in English **25**, 932 (1986).

[13] G. von Kiedrowski, in *Bioorganic chemistry frontiers* (Springer, 1993) pp. 113–146.

[14] N. Paul and G. F. Joyce, Current opinion in chemical biology **8**, 634 (2004).

[15] M. Eigen, Biophysical chemistry **85**, 101 (2000).

[16] D. T. Gillespie, The journal of physical chemistry **81**, 2340 (1977).

[17] D. T. Gillespie, Journal of computational physics **22**, 403 (1976).

[18] N. Vaidya, S. I. Walker, and N. Lehman, Chemistry & biology **20**, 241 (2013).

[19] P. Szabó, I. Scheuring, T. Czárán, and E. Szathmáry, Nature **420**, 340 (2002).

[20] N. Goldenfeld, *Lectures on phase transitions and the renormalization group* (Addison-Wesley, Advanced Book Program, Reading, 1992).

[21] J. P. Amend, D. E. LaRowe, T. M. McCollom, and E. L. Shock, Philosophical Transactions of the Royal Society of London B: Biological Sciences **368**, 20120255 (2013).

[22] C. Adami, "Information-theoretic considerations concerning the origin of life," (2014), arXiv:1409.0590.

[23] S. I. Walker and P. C. W. Davies, Journal of The Royal Society Interface **10** (2012), 10.1098/rsif.2012.0869.

[24] L. J. Mix, Astrobiology **15**, 15 (2015).

[25] G. F. Ellis, Interface Focus **2**, 126 (2012).

[26] S. J. Gould and N. Eldredge, Paleobiology **3**, 115 (1977).

[27] J. F. Atkins, R. F. Gesteland, and T. R. Cech, eds., *The RNA World, Third Edition* (Cold Spring Harbor, 2005).

[28] S. A. Benner, Origins of Life and Evolution of Biospheres **44**, 339 (2014).

[29] S. I. Walker, M. A. Grover, and N. V. Hud, PLoS One **7**, e34166 (2012).

[30] A. Chopra and C. H. Lineweaver, Astrobiology **16**, 7 (2016).

[31] J. von Neumann, *Theory of self-reproducing automata*, edited by A. W. Burks (University of Illinois Press, 1966).

[32] C. Marletto, Journal of The Royal Society Interface **12**, 20141226 (2015).

[33] S. Guiau, *Information Theory with New Applications* (McGraw-Hill Companies, 1977).

[34] I. Csiszar and J. Körner, *Information theory: coding theorems for discrete memoryless systems* (Cambridge University Press, 2011).

**SUPPLEMENTARY INFORMATION**

Herein, we explicitly measure the mutual information between two variables *as a timeseries variable* itself to track the progress of the phase transition from non-life to life. To generate a time series for mutual information we use the *pointwise mutual information*, P. Given two random variables $X = \{x_1, x_2...x_n\}$ and $Y = \{y_1, y_2....y_m\}$, P is quantified as:

$$\mathcal{P}(x_i : y_i) = \log \frac{p(x_i, y_i)}{p(x_i)p(y_i)}. \tag{2}$$

[33], where $p(x_i)$ and $p(y_i)$ are the probabilities of observing the event where $X$ is in state $x_i$ and $Y$ is in state $y_i$, respectively, and $p(x_i, y_i)$ is the joint probability of this event occuring. We generated probability distributions by counting the frequency of a given event (*e.g.* abundance of '0' and '1' monomers and of replicators of a given sequence composition) in our time series data. In the results presented here, the distributions were generated using time series data from an ensemble of 100 experimental runs over 10,000 time steps each. To ensure that the frequency based probability distributions were not biased by counting states from different phases of the system (see below), the frequencies were generated from data that sampled equally from both phases. The probabilities of different states therefore represent ensemble statistics which do not depend on time, while the particular ordering of states in a time series is used to determine the time series P. In stochastic systems, such as ours, P will fluctuate rapidly in time and is unlikely to yield useful insights. We therefore sum P over a fixed time window to yield the mutual information for that window. Explicitly, for a window size of $w$, the mutual information at time $t$ is defined as the average of $P(x_i : y_i)$, and is given by:

$$\mathcal{I}(X(t) : Y(t)) = \sum_{i=t-(w/2)}^{t+(w/2)} p(x_i, y_i) \log \frac{p(x_i, y_i)}{p(x_i)p(y_i)}. \tag{3}$$

[34]. This value will depend on time, not because the probabilities of different states will depend on time, but rather the realization of different states is time ordered. Determining an appropriate size for $w$ is important. If $w$ is too large, the entire measurement collapses into one value yielding no insights into how the system is evolving in time. By contrast, if $w$ is too small, fluctuations wash out interesting larger scale structure. We chose $w$

heuristically, such that the value of I(*R*;*E*), tracking the mutual information between replicators and environment, was relatively constant but large fluctuations could still be resolved. For the results presented *w* = 100Δ*t*, where Δ*t* = $0.1k_h^{-1}$ is the the resolution of the time series data in natural units. We note that different values of *w* change the results quantitatively, but not qualitatively: the system still maintains a non-zero value of the mutual information in the non-life phase which tends toward zero in the life phase.